\documentclass[a4paper]{article}
\usepackage{graphicx}
\usepackage{subfigure}
\usepackage{pstricks}
\usepackage{color}
\usepackage{amsfonts}
\usepackage{mathrsfs,amsmath}
\usepackage{epsfig}
\usepackage{amsmath,amssymb,amsthm,graphicx,latexsym}
\renewcommand{\arraystretch}{2}

\newcommand{\be}{\begin{equation}}
\newcommand{\ee}{\end{equation}}
\newcommand{\bea}{\begin{eqnarray}}
\newcommand{\eea}{\end{eqnarray}}
\newcommand{\bml}{\begin{subequations}}
\newcommand{\eml}{\end{subequations}}
\newcommand{\bfig}{\begin{figure}}
\newcommand{\efig}{\end{figure}}

\newcommand{\slashed}{\hspace{-1.1ex}/}

\begin{document}
\title{\LARGE \textsc{\fontsize{18}{22}\selectfont \sffamily \bfseries{From Extended theories of Gravity to Dark Matter }}}
\author{{\bf\small $^1$Sayantan Choudhury\footnote{\tiny sayantan@theory.tifr.res.in},  $^1$\small Manibrata Sen\footnote{\tiny manibrata@theory.tifr.res.in} ,
$^2$\small Soumya Sadhukhan\footnote{\tiny soumyasad@imsc.res.in }}\\
\small{$^1$Department of Theoretical Physics, Tata Institute of Fundamental Research, Colaba, Mumbai - 400005, India.}\\
\small{$^2$Institute of Mathematical Sciences, C.I.T Campus, Taramani, Chennai - 600113, India.} } 
\date{\vspace{-5ex}}
\maketitle
\begin{abstract}
\small In this work, we propose different models of extended theories of gravity, which are minimally coupled to the SM fields, 
	to explain the possibility of a dark matter (DM) candidate, without ad-hoc additions to the Standard Model (SM). We modify the gravity sector by allowing
	 quantum corrections motivated from local $f(R)$ gravity, and non-minimally coupled gravity with SM sector and dilaton field. Using an effective field theory (EFT) framework,
	 we constrain the scale of the EFT and DM mass. We consider two cases-Light DM (LDM) and Heavy DM (HDM),
	 and deduce upper bounds on the  DM annihilation cross section to SM particles.\\
	 \begin{flushright}
	 {\textcolor{red}{ \bf TIFR/TH/16-17}}
	 \end{flushright}
\end{abstract}

\bigskip

        Majority
     of the matter in this universe occurs in the form of 
     ``dark matter''(DM). A general approach to explain the existence of a DM candidate is an ad-hoc addition of a
     new particle, intended to serve as the DM. However, that is not always theoretically well-motivated.
         To explain the genesis of a DM candidate, we propose an alternate framework based on the principles of EFT.

         Here, we start with the extended version of gravity sector keeping the SM matter sector unchanged. The cases considered are given by:
	\be\begin{array}{lll}\label{eq321}
		\displaystyle   S=\left\{\begin{array}{lll}
			\displaystyle  
			\int d^{4}x\sqrt{-g}\left[\frac{\Lambda^2_{UV}}{2}\left(aR+bR^{n}\right)+{\cal L}_{SM}\right]\,,~~~~~~ &
			\mbox{ $f(R)$ gravity}  \\ 
			\displaystyle   
			\int d^{4}x\sqrt{-g}\left[\frac{\Lambda^2_{UV}}{2}\left(1+\xi \frac{\phi^2}{\Lambda^2_{UV}}\right)R+{\cal L}[\phi]+{\cal L}_{SM}\right]\nonumber\,~~~~~~ & 
 			\mbox{ non-minimal gravity}  
		\end{array}
		\right.
	\end{array}\ee 
	where  $\Lambda_{UV}$ is the scale of new physics. For these classes of gravity theory, after conformally transforming the metric in the {\it Jordan frame} into {\it Einstein frame} we get:
	\bea S\Longrightarrow \tilde{S}&=&\int d^{4}x\sqrt{-\tilde{g}}\left[\frac{\Lambda^2_{UV}}{2}\tilde{R}-\frac{1}{2}\tilde{g}^{\mu\nu}\partial_{\mu}\phi\partial_{\nu}\phi-V(\phi)+ e^{-\frac{2\sqrt{2}}{\sqrt{3}}\frac{\phi}{\Lambda_{UV}}}{\cal L}_{SM}\right], \eea
	where the new scalar field, ``dilaton'' $(\phi)$, generated through conformal transformations, is identified as the DM candidate. The potential term $V(\phi)$ has terms which involve self interaction of 
	DM and can be used to constrain the parameters of these extended theories, as shown in \cite{Sen}.
  Assuming scale of new physics is large, we expand Lagrangian as
     \begin{equation}\label{frac123}
	 e^{-\frac{\phi}{\Lambda_{UV}}}{\cal L}_{SM}\xrightarrow{{\cal Z}_2}\left\{1+ \frac{\phi^2}{2\Lambda_{UV}^2} \right\}{\cal L}_{SM}
     \end{equation}
     Thus, here DM couples {\it universally} to SM, at the higher orders. We consider the cases of LDM $(10~{\rm GeV}<M<350~{\rm GeV})$ and HDM $(350~{\rm GeV}<M<1~{\rm TeV})$. The scale
     associated with HDM is larger than that of LDM and this distinguishes the two cases.
     We calculate the relic density using (upto $p$ wave expansion) \cite{Kolb}
     \begin{equation}
      \left[ \sigma v\right]_{NR}=a(\Lambda_{UV},M) +b(\Lambda_{UV},M) v^2+\cdots .\nonumber
     \end{equation}
     
    Using the experimental bounds on relic density observed by Planck 2015 \cite{Planck},
     we deduce constraints on the DM mass and scale of new physics $(M,\Lambda_{UV})$.
 \begin{figure}[!h]
           \includegraphics[width=4cm]{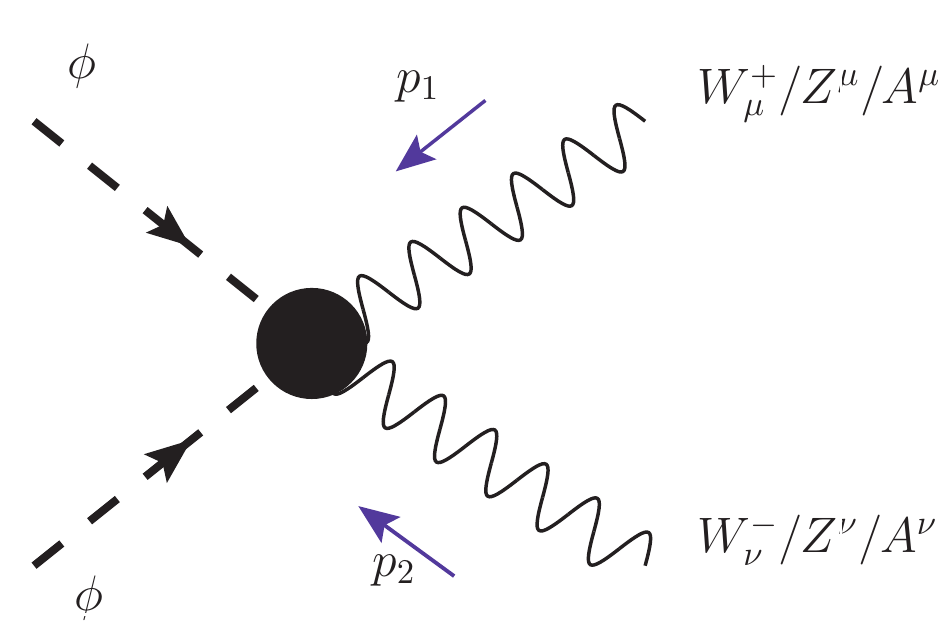}
           \includegraphics[width=4cm]{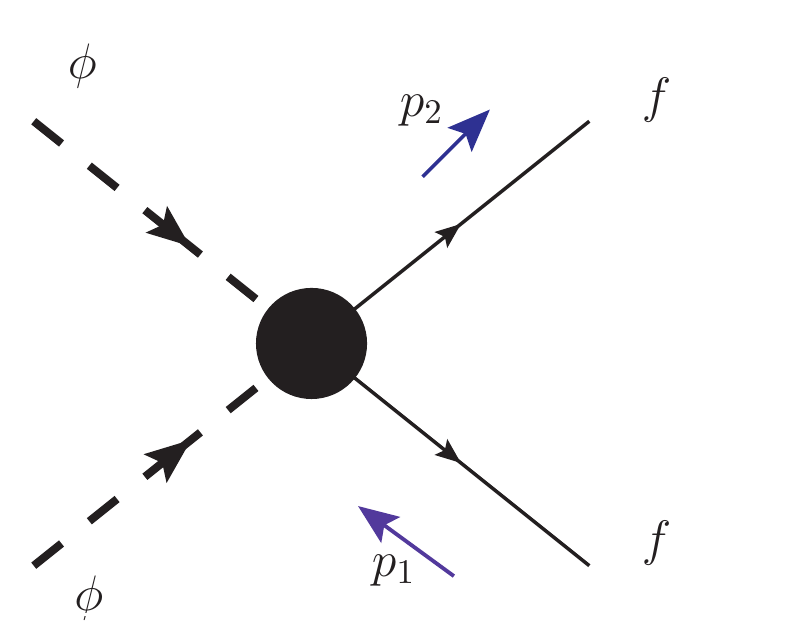}
           \includegraphics[width=4cm]{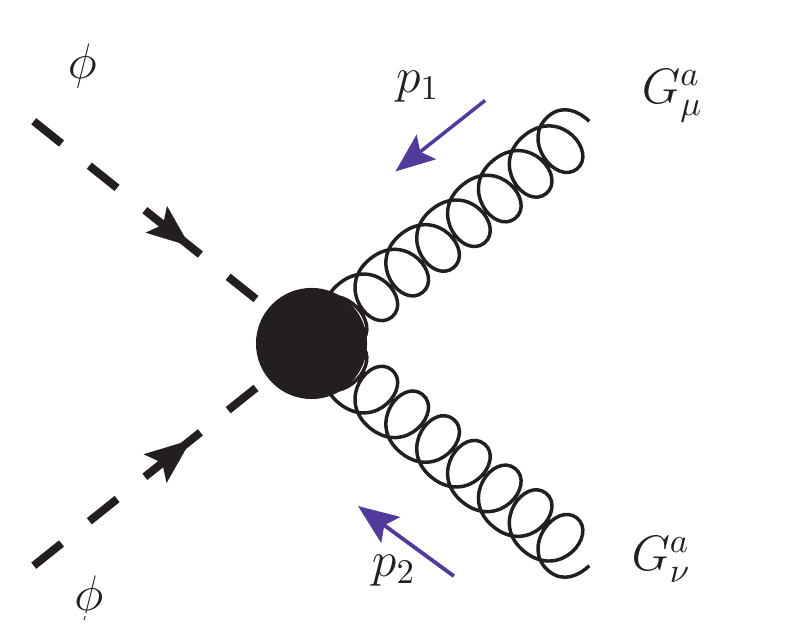}
           \includegraphics[width=4cm]{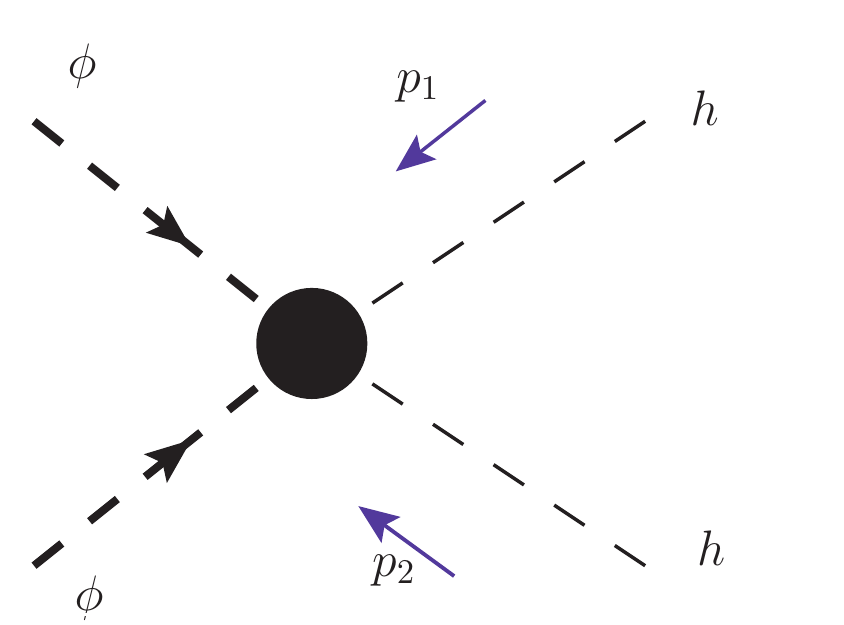}
           \caption{\small Effective processes contributing to relic density of DM. Here $2\rightarrow 3$ and $2\rightarrow 4$ processes are suppressed.}
           \label{fig1}
\end{figure}
 
\begin{figure}[!htb]
\includegraphics[width=8.4cm,height=4cm] {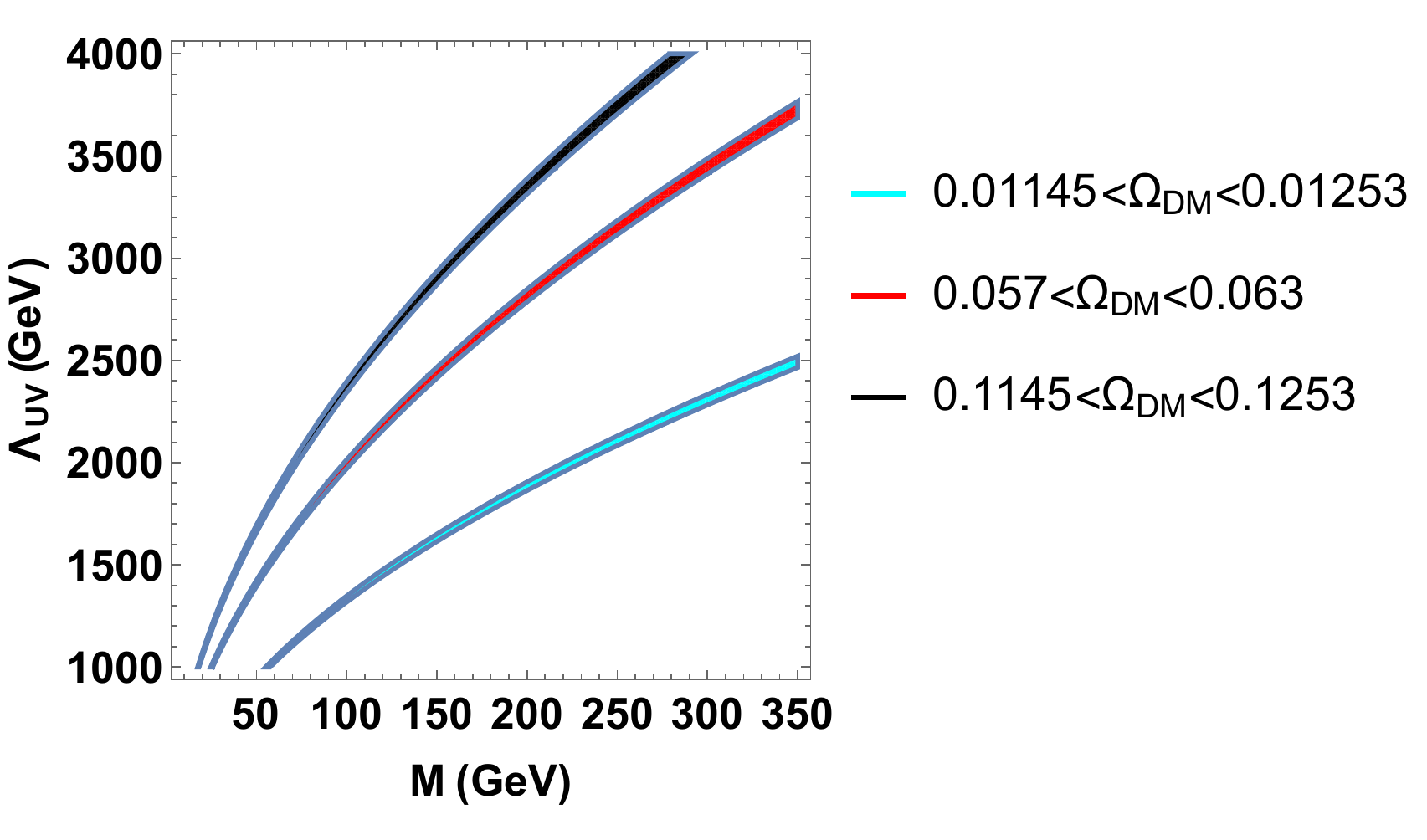}
\includegraphics[width=7.4cm,height=4cm] {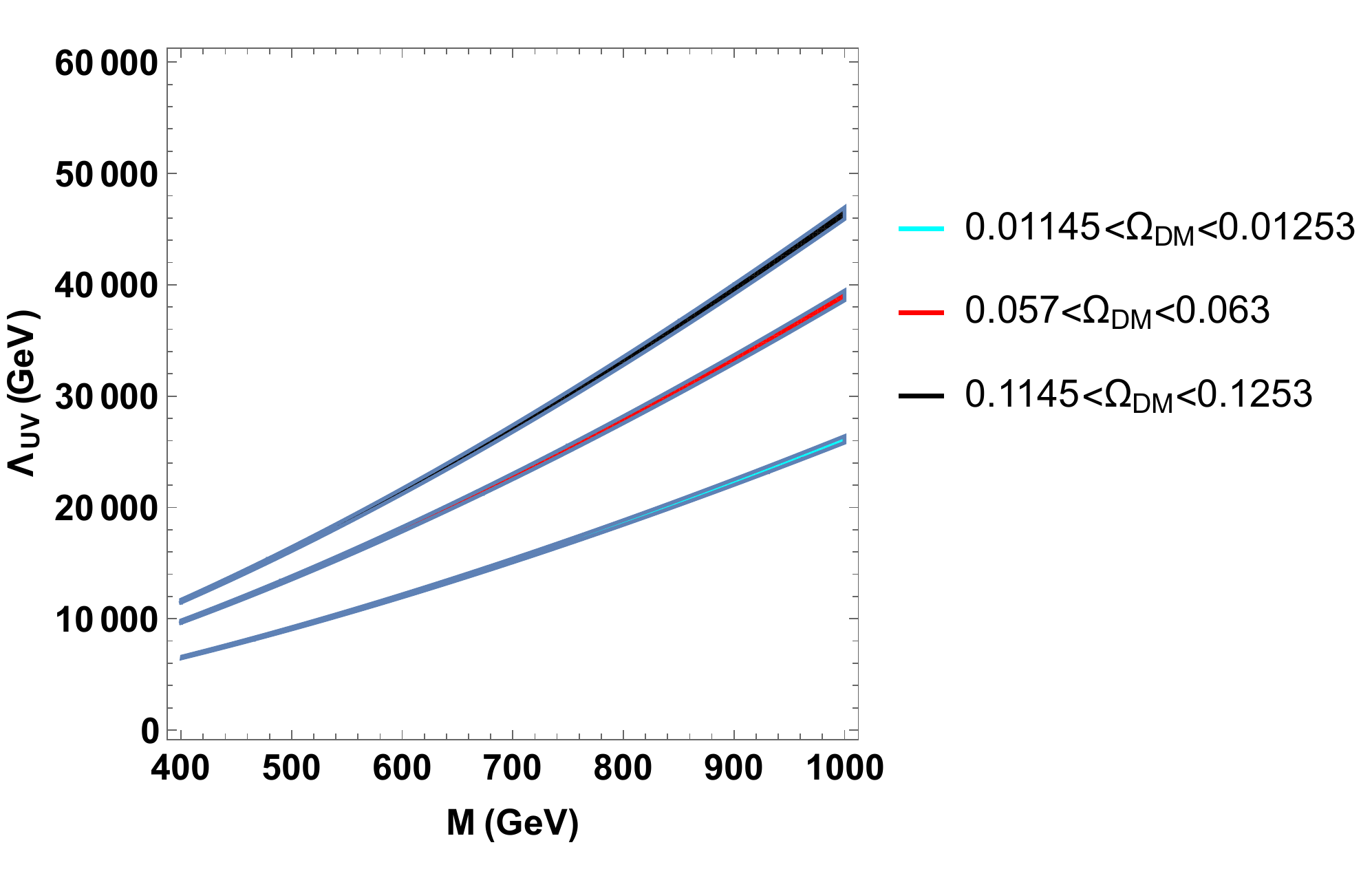}
\caption{\small Allowed parameter space for LDM (left) and HDM (right) candidate. Black denotes $2\sigma$ bounds, while red and cyan 
denotes $50\%$ and $10\%$ of the $2\sigma$  relic bounds.}
  \label{fig2}
\end{figure}
\begin{figure}[!htb]
\begin{center}
\includegraphics[width=6.4cm,height=2.6cm]{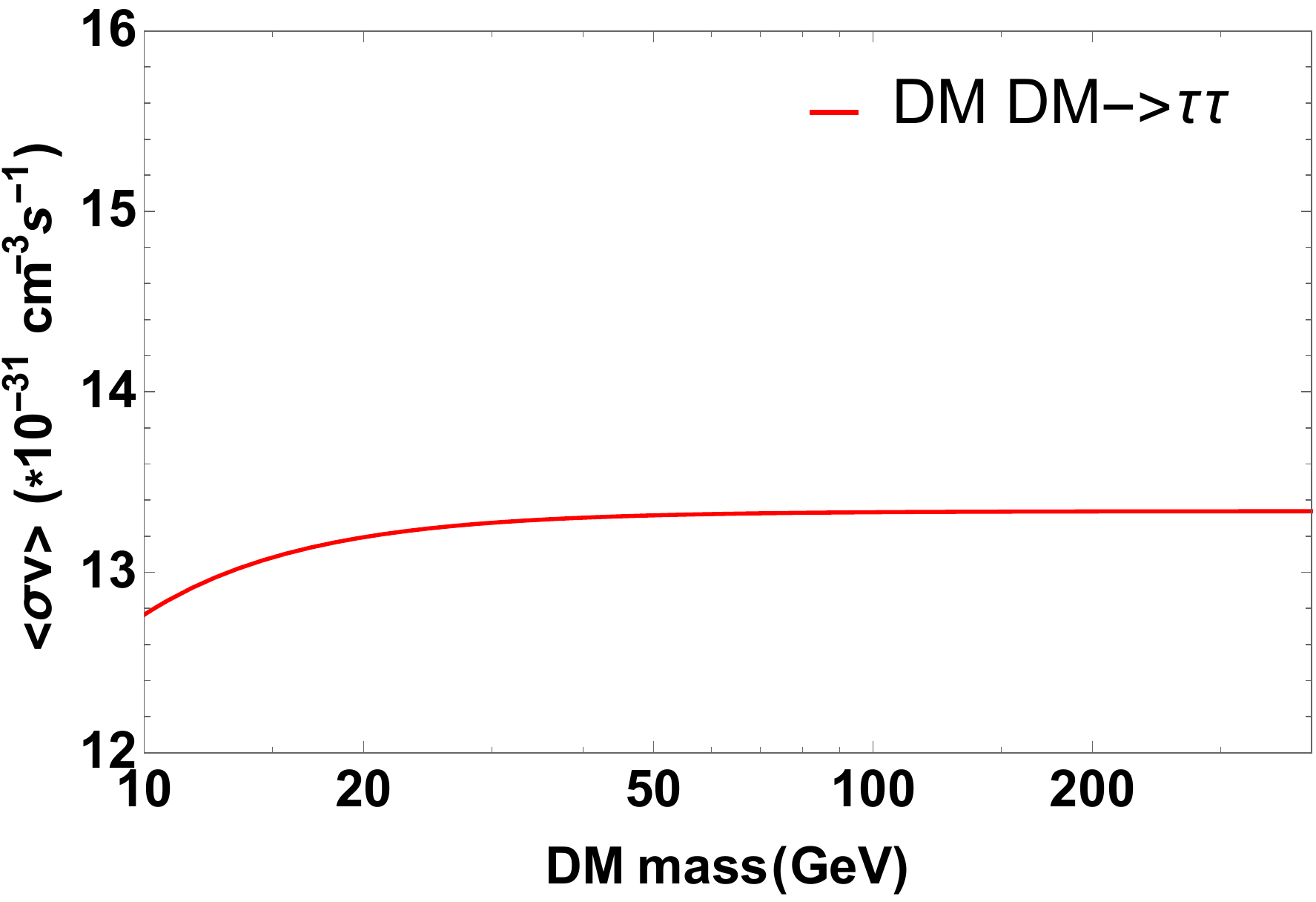}
\includegraphics[width=6.4cm,height=2.6cm]{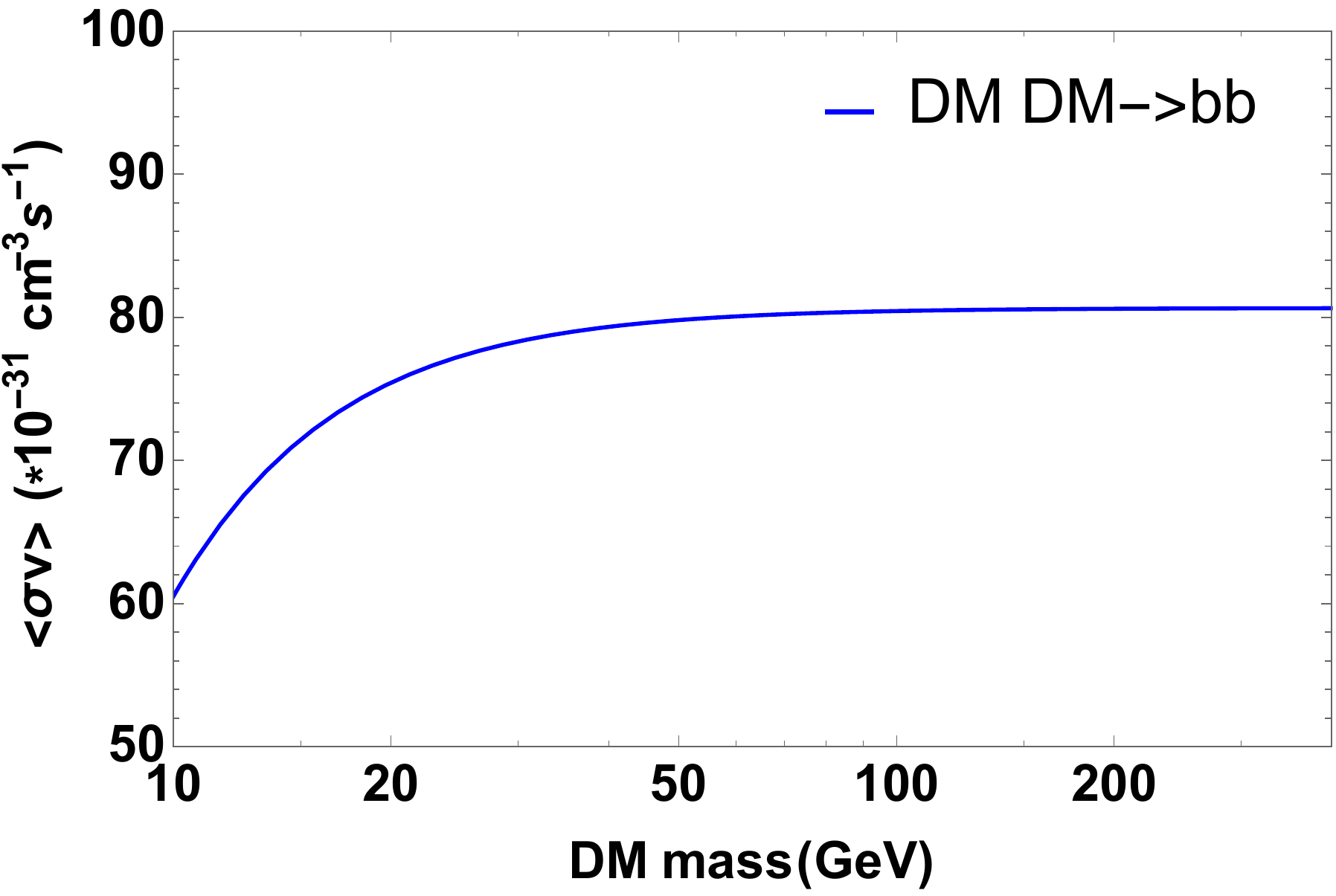}
\includegraphics[width=6.4cm,height=2.6cm]{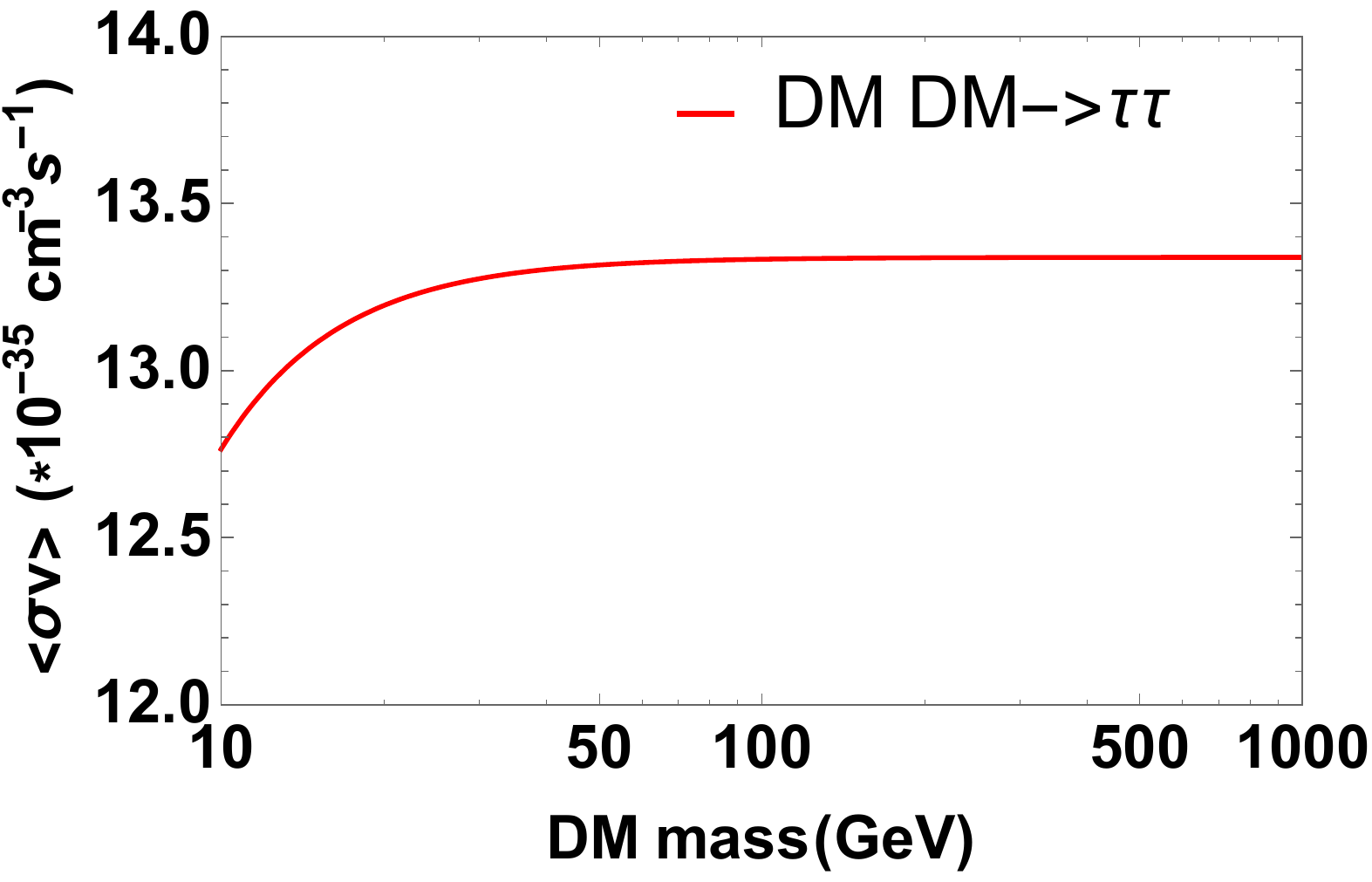}
\includegraphics[width=6.4cm,height=2.6cm]{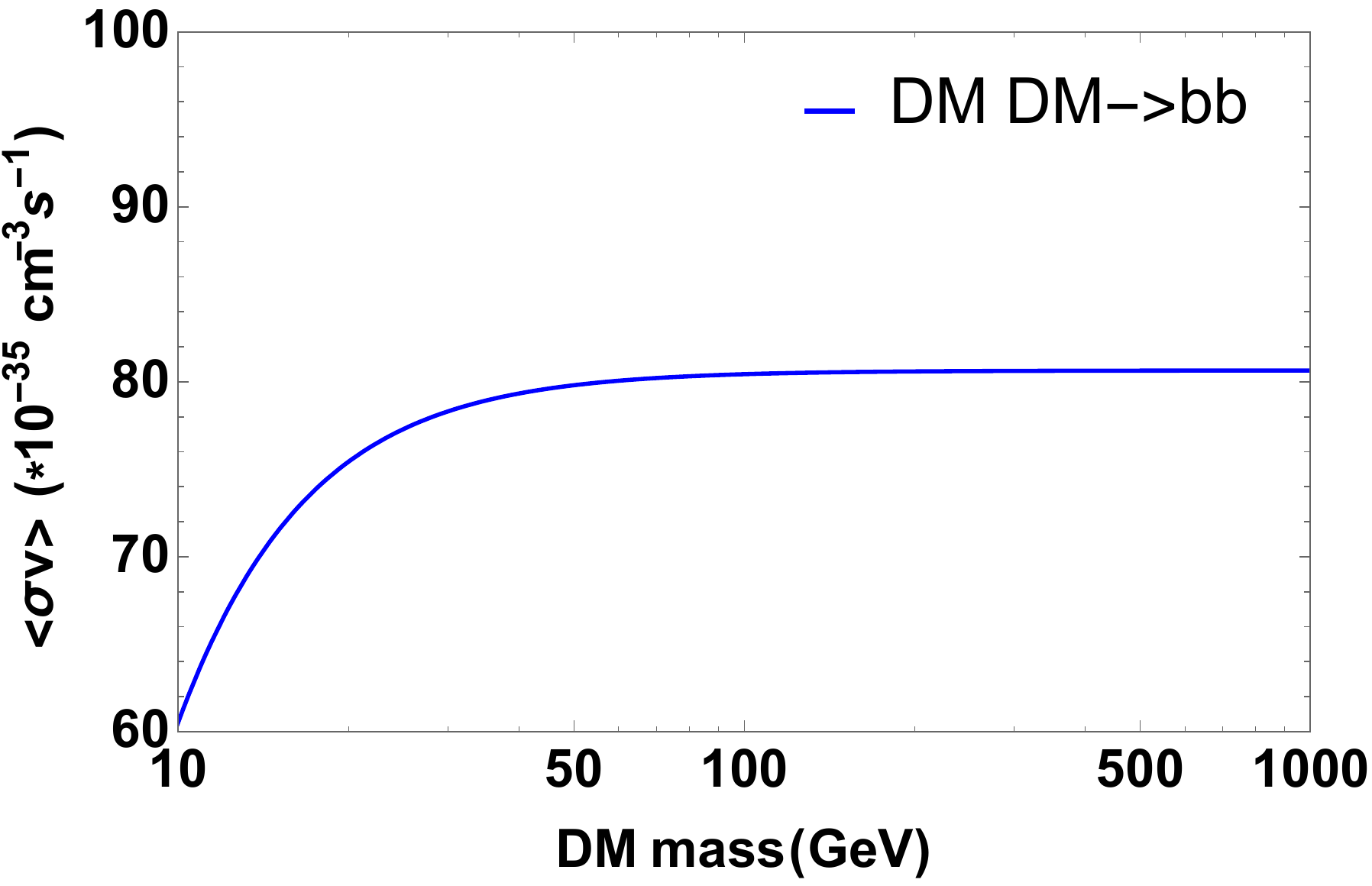}
\end{center}
 \caption{\small Annihilation cross-sections of DM to $b~\bar{b}$ and $\tau~\bar{\tau}$ for LDM (top panel) and HDM (bottom panel) candidate.
The values obtained are well within the bounds given in \cite{Fermi}.}
\label{fig3}
\end{figure}

We present plots of the allowed parameter space in Fig.~\ref{fig2}. We have allowed the fact that the dilaton may not be the only DM candidate, in which case it can contribute
to a certain fraction of the relic. We show cases where the dilaton contributes to $10\%$ and $50\%$ of the $2\sigma$  relic bounds.
In Fig.~\ref{fig3}, we show the annihilation cross section of DM to $b~\bar{b}$ and $\tau~\bar{\tau}$, assuming these
to be the main annihilation channels. We see similar features for both
LDM and HDM candidates. However, for HDM, the  cross-sections are
well below the current experimental sensitivity, and cannot be probed by present experiments.
The cases for other channels are presented in \cite{Sen}.
           
 Therefore, in conclusion, we have presented a DM candidate which is generated solely from the
	  gravity sector. This shows that although we start with an unchanged matter sector, extended theories gravity automatically allow for a
	  dark matter candidate. Based on 
	  whether our candidate is LDM or HDM, we have presensed the corresponding allowed parameter space and bounds.

\section*{\small Acknowledgements}
\small SC would like to thank DTP, TIFR, Mumbai for providing the Visiting Fellowship. MS would like to thank Amol Dighe and Basudeb Dasgupta for useful discussions.

{}

\end{document}